\documentclass[aps,prl,preprint,unsortedaddress]{revtex4-1}
\usepackage[dvips]{graphicx}

\begin{document}

\title{Comment on\\
       ``Logarithmic Oscillators: Ideal Hamiltonian Thermostats''}

\author{M. Mel\'endez}
\affiliation{Dpto. F\'isica Fundamental, UNED\\
             Senda del Rey, 9\\
             28040 Madrid} 
\email{mmelendez@fisfun.uned.es}

\author{Wm. G. Hoover}
\affiliation{Ruby Valley Research Institute\\
             Highway Contract 60 Box 601\\
             Ruby Valley 89833-9803 Nevada}

\author{P. Espa\~nol}
\affiliation{Dpto. F\'isica Fundamental, UNED\\
             Senda del Rey, 9\\
             28040 Madrid}

\date{\today}

\begin{abstract}
Campisi, Zhan, Talkner and H\"anggi have recently proposed a novel Hamiltonian
thermostat which they claim may be used both in simulations and experiments. We
show, however, that this is not possible due to the length and time scales
involved, which depend \textit{exponentially} on the total energy of the system.
The implementation suggested by Campisi \textit{et alii} implies equilibration
times greater than the age of the universe for systems with more than a few
dozen particles.
\end{abstract}

\maketitle

Campisi, Zhan, Talkner and H\"anggi have recently proposed the use of the
logarithmic oscillator as a novel Hamiltonian thermostat \cite{Campisi} and have
claimed that ``it may be implemented not only in a computer but also with
real-world experiments''. Unfortunately, this is not possible in most
practical applications, because of the length and time scales involved, which
depend \textit{exponentially} on the total energy of the system (which in turn
should be set to a large value, according to \cite{Campisi}).

Consider the implementations suggested in \cite{Campisi}: an ion in a
two-dimensional Coulomb potential generated by a thin oppositely charged wire
(or a laser  beam) that interacts through short ranged forces with a gas of
neutral atoms confined in a box. The size $L$ of the box is chosen  according to
     $L=2\sigma\sqrt{e^{2\beta E_{\rm tot}}-1}$,
which guarantees that the ion never leaves the box. In addition, the energy
$E_{\rm  tot}$ should  be ``large'' in order to reduce the effect of an
approximation introduced in the logarithmic potential (for $2$ and $3$
particles, the values chosen in \cite{Campisi} were $5 k_BT$ and $8 k_BT$,
respectively). Furthermore, Campisi \textit{et alii} estimate that, if the
number $N$ of atoms increases, then $E_{tot}$ should increase accordingly, with
$E_{\rm tot}\propto 3Nk_BT/2$.  For a system  of just $26$ atoms this means that
the length $L$ of the box is larger than the diameter of planet Earth (setting
$\sigma = 10^{-10}$ m). But, apart from the obvious problems that such a setup
would imply, the system formed by the ion plus $26$ atoms would take
extraordinarily long to equilibrate. A crude estimate of the mean free time for
the logarithmic oscillator gives
\[
\tau \sim \sqrt{\frac{m}{k_BT}} \frac{L^2 \sigma}{4\pi  N \sigma^2}
     \sim  10^{19} \ {\rm s},
\]
($m=1$ amu, $T=1$ K), exceeding the age of the universe.

The same problems arise in computational simulations. Even in  simulations with
only two or three particles, the number of time steps necessary to sample the
theoretical canonical distribution is of the order of $10^9$ steps if one wishes
a reasonable reproduction of the results presented in \cite{Campisi}.

Campisi \textit{et alii} have also suggested that the logarithmic oscillator can
be used in other settings as, for example, ``to study the response of a system
to a varying temperature''. Following this suggestion, we have explored the
behaviour of \textit{two} logarithmic oscillators at different temperatures
brought into contact by means of two $9$-atom $\phi^4$ chains
\cite{Aoki-Kusnezov} with periodic boundary conditions. No tendency towards the
theoretical linear temperature gradient was observed.

The logarithmic oscillator Hamiltonian presented by Campisi \textit{et alii}
is undoubtedly the simplest example for which Gibbs's statistical mechanics
implies the canonical distribution in phase space, but although log-thermostats
have excellent pedagogical values, they are not useful in most practical
applications, whether simulations or experiments.


\section{Addendum of 17 June 2012 (\textit{Revised 18 July 2012})}

\subsection{$\phi^4$ Chains with Two Log-Thermostat Particles}

The comprehensive investigation of the ``$\phi^4$'' atomistic model for heat
flow carried out by Aoki and
Kusnezov \cite{Aoki-Kusnezov,Aoki-Kusnezov-2,Aoki-Kusnezov-3} showed that this
model behaves ``normally'', even in one space dimension.  Heat flows through a
chain of $\phi^4$ particles according to Fourier's Law \cite{Hoover1},
$Q_x = -\kappa (dT/dx)$.
Thus this model can provide good test cases for the logarithmic oscillator
thermostat.  The model we investigate here is a periodic chain of 20
one-dimensional particles.  Two are ``log-thermostat'' particles, characterized
by their individual specified ``thermostat temperatures'' $\{T\}$, and
interacting with their lattice sites $\{q_0\}$ with a logarithmic potential:
\[
\phi_{\rm log} = (T/2)\ln ( \delta q^2 + 0.1) \ ; \ \delta q = q - q_0 \ .
\]
The remaining eighteen are $\phi^4$ particles, tethered to their lattice sites
$\{q_0\}$ with a \textit{quartic} potential :
\[
\phi_{\rm tether} = (1/4)(q-q_0)^4 \ .
\]
In addition to these two types of lattice-site potentials all 20
nearest-neighbor pairs interact with a Hooke's-Law potential,
\[
\phi(q_i,q_{i+1}) = (\kappa /2)(|q_i - q_{i+1}| - 1 )^2 \ ;
\ \kappa = 1.00 \ {\rm or} \ 0.10 \ .
\]
Because the log-thermostat model is imagined to be ``weakly coupled'' to the
chain we considered a model with a much smaller force constant $\kappa = 0.1$
linking the two thermostat particles to their four neighbors in the chain.
Experiments with $\kappa = 0.01$ showed no tendency at all towards equilibration
with simulations of $10^9$ time steps.  With initial velocities $\pm 1$
alternating along the chain the longtime averaged temperatures along the chain
reflect the initial conditions rather than the thermostat temperatures, ending
up with all the time-averaged kinetic temperatures near
$\langle \ p^2 \ \rangle = 0.5$ .
The log-thermostats are apparently unable to absorb much energy in a reasonable
time \cite{Melendez}.

\begin{figure}
  \includegraphics[width=0.5\textwidth,angle=-90]{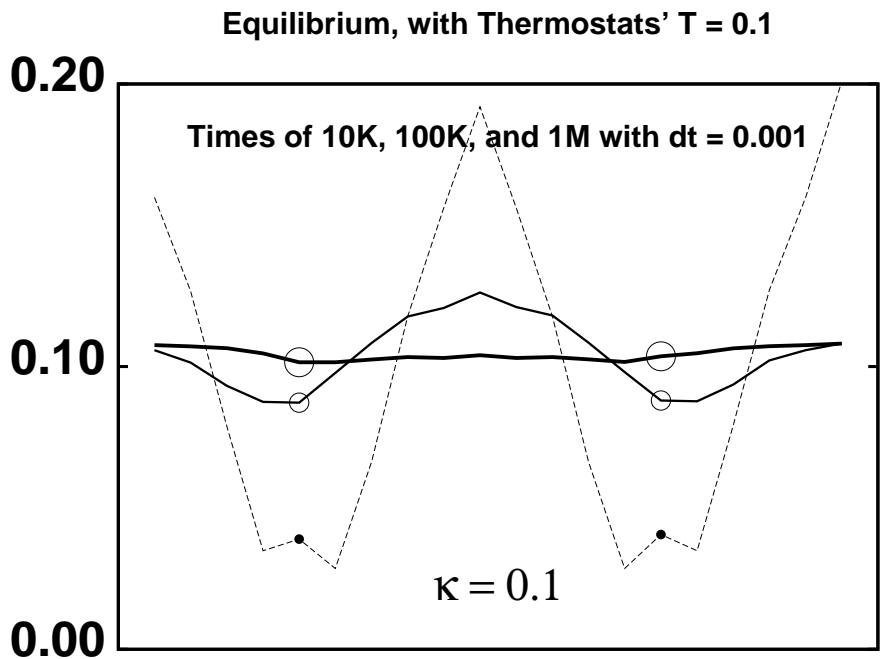}
  \caption{\label{eq}Equilibrium temperature profiles for a 20-particle periodic
    chain.\vspace{10mm}}
\end{figure}

Equilibrium simulations with alternating initial velocities $\pm \sqrt{0.2}$
along the chain and with both specified thermostat temperatures equal to $0.10$
were more nearly successful.  Figure \ref{eq} shows that the time-averaged
kinetic temperatures along the chain are within 8\% of the specified temperature
$0.10$ after a simulation of $10^9$ time steps, corresponding to a time of one
million in reduced units. Evidently, under propitious conditions log-thermostat
temperature control \textit{can} approach equilibrium on a sufficiently long
timescale.

\begin{figure}
  \includegraphics[width=0.5\textwidth,angle=-90]{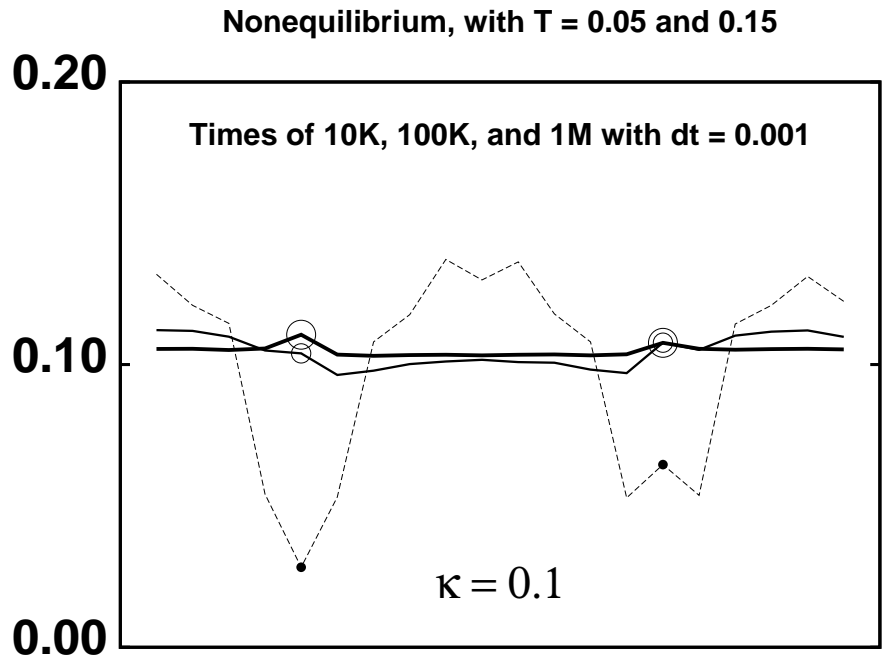}
  \caption{\label{neq}``Nonequilibrium'' temperature profiles for a 20-particle
    chain.}
\end{figure}

We next carried out a similar, but \textit{nonequilibrium} simulation, with the
same initial conditions but with different specified thermostat temperatures:
$0.05$ for thermostat Particle 5 and $0.15$ for thermostat Particle 15. The
temperature profile which resulted (again with $10^9$ fourth-order Runge-Kutta
time steps) was scarcely different to the equilibrium one (see figure
\ref{neq}). The log-thermostats were \textit{unable} to provide a nonequilibrium
temperature profile.

But why do logarithmic thermostats fail? Apart from the large time intervals
mentioned above \cite{Melendez}, there is another more fundamental reason for
their failure in \textit{nonequilibrium} problems, traceable to their
Hamiltonian heritage \cite{Hoover2}: Deterministic nonequilibrium heat-flow
problems generate \textit{fractal} phase-space distributions, with a vanishing
phase volume.  A Hamiltonian system obeying Liouville's Theorem in phase space,
$df(q,p)/dt \equiv 0$ , simply \textit{cannot} produce a fractal.

Aoki and Kusnezov showed that heat flow through a $\phi^4$ chain generates
fractal phase-space distributions, with a dimensionality reduced from the
equilibrium Gibbs'
distribution \cite{Aoki-Kusnezov,Aoki-Kusnezov-2,Aoki-Kusnezov-3}.  Hoover
\textit{et alii} \cite{Hoover-Aoki} showed that similar fractals result using
seven different thermostat types (none of which obeys the equilibrium version of
Liouville's Theorem).

\newpage

\subsection{Lennard-Jones Potentials}

\begin{figure}
  \includegraphics[width=0.5\textwidth,angle=-90]{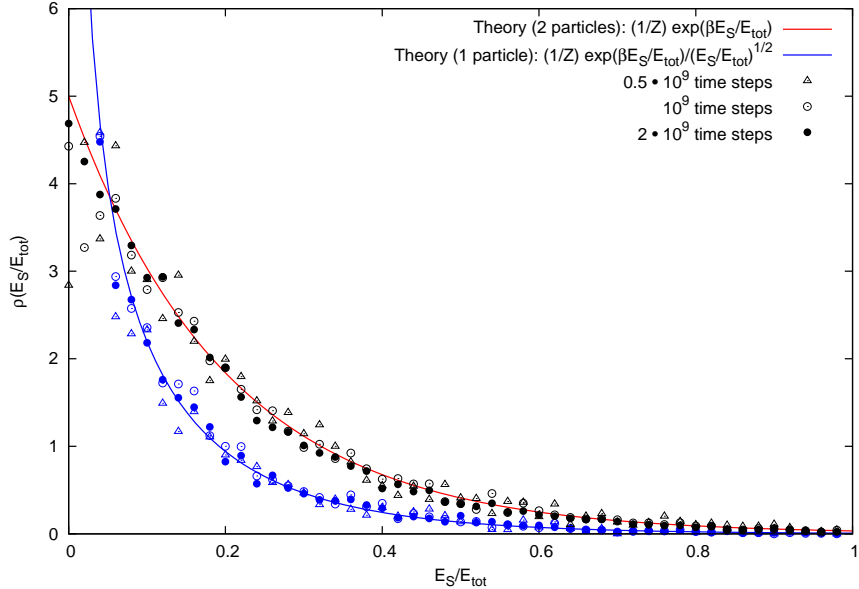}
  \caption{\label{densities}
    Probability distribution for $E_S/E_{tot}$ in the original numerical
    experiment proposed by Campisi \textit{et alii} in \cite{Campisi}. $E_S$ is
    the energy of a system interacting with the logarithmic oscillator. The
    theoretical prediction follows the solid line (red for two particles and
    blue for only one). The black points correspond to the numerical results for
    $t = 0.5\cdot 10^6, 10^6$ and $2 \cdot 10^6$ for a system of two particles,
    as in the original article (the time step was set to $\Delta t = 0.001$).
    The blue points correspond to a system of only one thermostated particle,
    which also takes about the same time to converge to the prediction.}
\end{figure}

Even the original one-dimensional simulations proposed by Campisi \textit{et
alii} for a couple of particles turn out to imply very long simulation times. We
carried out simulations with one and two particles setting the mass of the
logarithmic oscillator equal to ten particle masses. A classic fourth-order
Runge-Kutta integrator took $2 \cdot 10^9$ time steps to generate a reasonable
reproduction of the energy histogram presented in \cite{Campisi} (see figure
\ref{densities}).

In their discussion of a three-dimensional simulation, Campisi \textit{et alii}
pointed out that an increase in the number of particles led to a very
significant departure from the predicted velocity distribution. The solution
suggested was simply to increase the total energy of the system by
$\Delta E \propto 3Nk_BT/2$. This solution, however, leads to the exponential
increase in the typical lengths and times for the logarithmic oscillator that we
have already explained above.

Our investigations reveal that unless the initial conditions and the problem are
carefully ``tuned'' the thermostat is ineffective at equilibrium, even for
extraordinarily long simulation times.  The situation away from equilibrium is
worse yet, as the thermostat fails to act rapidly enough to affect change.
We conclude that log-thermostats are not useful in most practical applications,
whether simulations or experiments.

We would like to thank Campisi \textit{et alii} for correcting a mistake in the
previous version of this article \cite{Campisi-Reply}.


\section{Addendum 29 January 2013}

Our comment was published in Physical Review Letters on the 11$^{\rm th}$ January
2013 \cite{PRL-Comment}, followed by a reply \cite{PRL-Reply} where Campisi and
his colleagues proposed a new experimental arrangement for the logarithmic
oscillator, without the unreasonable time or length scales that we had
described. The number of degrees of freedom in the original experiment was
reduced to one third by forcing the neutral atoms and logarithmic oscillator ion
to move along a single dimension. Table \ref{Campisi-table}, taken from the
reply, illustrates the exponential growth of mean free times $\tau$ and box
lengths $L$ as the required precision $H_{KS}$ or the number of particles $N$
increase.

Campisi \textit{et alii} claimed that this version of the experiment could be
implemented with present day cold-atom technology \cite{Bloch}. Having no prior
experience with cold-atom physics, we contacted Prof. I. Bloch, who kindly lent
us some of his time and confirmed that such a precise one-dimensional setup,
though ``challenging'', should be feasible in principle. We are grateful for his
helpful comments.

Although the magnitudes shown in the table are correct, they are slightly
misleading because they assume that the system of interest begins at (or very
near) the ``thermostat temperature''. However, if we assume that the initial
temperature is off by $\Delta T$ degrees, then the logarithmic oscillator will
have to absorb at least $\Delta E = Nk_B\Delta T/2$ units of energy. For $N=20$
and $\Delta T = 5 \mathrm{K}$, for example, the energy absorbed must be about
$\Delta E = 50 k_B$. Compare this value to those in the table, where the total
energy of system plus oscillator never exceeds $30k_B$.

\begin{table}
  \caption{Total energy, box lengths and mean free times for the logarithmic
    oscillator experiment as a function of the number of degrees of freedom,
    $N$, and the required precision, $H_{KS}$, measured as a Kolmogorov-Smirnov
    distance (from Campisi \textit{et alii} \cite{PRL-Reply}).}
  \centering
  \begin{tabular}{c c c c c}
  \hline\hline
  $N$ & $H_{KS}$ & $E_{tot}/k_B$ & $L\ \mathrm[m]$ & $\tau\ \mathrm[s]$ \\
  \hline
  20 & 0.005 & 16.45 & $3\times10^{-1}$ & $1\times10^{-3}$ \\
  20 & 0.01  & 14.8 & $5\times10^{-2}$ & $3\times10^{-4}$ \\
  20 & 0.02  & 13.1 & $9\times10^{-3}$ & $5\times10^{-5}$ \\
  \hline
  30 & 0.02  & 18.1 & $1\times10^{0}$ & $5\times10^{-3}$ \\
  40 & 0.02  & 23.1 & $2\times10^{2}$ & $5\times10^{-1}$ \\
  50 & 0.02  & 28   & $3\times10^{4}$ & $6\times10^{1}$ \\
  \hline\hline
  \end{tabular}
  \label{Campisi-table}
\end{table}

Logarithmic oscillators indeed ``possess an infinite heat capacity'', but this
statement is easily misunderstood. The logarithmic oscillator's mean kinetic
temperature is not a function of its energy (if one considers time averages
with intervals that are very large compared to the period of oscillation).
In practice, though, a logarithmic oscillator \textit{cannot} absorb an
arbitrary amount of heat because any physical potential will lack a singularity
at the origin and the size of the experiment, $L$, will limit the amount of
energy that the oscillator may absorb, so that
\[\Delta E_{max.} = \frac{1}{2}k_BT\ \ln \left(\frac{L^2 + b^2}{b^2}\right),\]
which is an extremely slowly growing function of $L$.

Our comment pointed out that applying \textit{two} logarithmic oscillators, with
different temperatures, to a chaotic Hamiltonian system failed to create the
expected linear temperature gradient.  In their Reply, Campisi \textit{et alii}
disregarded this observation, arguing that their Letter suggested temperatures
that varied in time and not in space, so that our simulations were not relevant
to their work.  This conclusion strikes us as ill-conceived. Unless they can
somehow explain how to change a system's temperature \textit{homogeneously}, one
would expect to find that a \textit{time}-varying temperature would necessarily
create gradients in \textit{space}.

Consequently we stand by our claim that the logarithmic oscillator cannot be 
used an an effective thermostat in practical applications.


\end{document}